\documentclass{ws-ijmpe}
\usepackage{xcolor}
\usepackage[super,compress]{cite}
\usepackage{graphicx}
\usepackage{epsfig}
\usepackage{psfrag}
\usepackage{braket}
\begin{document}

\markboth{M. M. Saez, K. J. Fushimi, M. E. Mosquera, O. Civitarese}{Limits on active-sterile neutrino mixing parameters using heavy nuclei abundances}

\catchline{}{}{}{}{}

\title{Limits on active-sterile neutrino mixing parameters using heavy nuclei abundances}

\author{M. M. Saez}

\address{Facultad de Ciencias Astron\'omicas y Geof\'{\i}sicas, University of La Plata. Paseo del Bosque S/N\\
1900, La Plata, Argentina.\\
msaez@fcaglp.unlp.edu.ar}

\author{K. J. Fushimi}

\address{Facultad de Ciencias Astron\'omicas y Geof\'{\i}sicas, University of La Plata. Paseo del Bosque S/N\\
1900, La Plata, Argentina.\\
kfushimi@fcaglp.unlp.edu.ar}

\author{M. E. Mosquera}

\address{Dept. of Physics, University of La Plata, c.c.~67\\
Facultad de Ciencias Astron\'omicas y Geof\'{\i}sicas, University of La Plata. Paseo del Bosque S/N\\
 1900, La Plata, Argentina\\
mmosquera@fcaglp.unlp.edu.ar
}

\author{O. Civitarese}

\address{Dept. of Physics, University of La Plata, c.c.~67\\
 1900, La Plata, Argentina\\
osvaldo.civitarese@fisica.unlp.edu.ar}
\maketitle

\begin{history}
\received{Day Month Year}
\revised{Day Month Year}
\end{history}

\abstract {The production of heavy-mass elements due to the rapid
neutron-capture mechanism (r-process) is associated with
astrophysical scenarios, such as supernovae and neutron-star
mergers. In the r-process the capture of neutrons is followed by
$\beta$-decays until nuclear stability is reached. A key element in
the chain of nuclear weak-decays leading to the production of
isotopes may be the change of the parameters controlling the
neutrino sector, due to the mixing of active and sterile species. In
this work we have addressed this question and calculated
$\beta$-decay rates for the nuclei involved in the r-process chains
as a function of the neutrino mixing parameters. These rates were
then used in the calculation of the abundance of the heavy elements
produced in core-collapse supernova and in neutron-star mergers,
starting from different initial mass-fraction distributions. The
analysis shows that the core-collapse supernova environment
contributes with approximately $30\%$ of the total heavy nuclei
abundance while the neutron-star merger contributes with about
$70\%$ of it. Using available experimental data we have performed a
statistical analysis to set limits on the active-sterile neutrino
mixing angle and found a best-fit value $\sin^2 2\theta_{14}=0.22$, a value comparable with those found in other studies reported in the literature.}

\ccode{PACS numbers:}

\maketitle

\section{Introduction}

The production of nuclei in the Universe is directly related to weak
interactions between
 neutrinos and matter. Neutrino interactions can exchange protons and
 neutrons, therefore they affect the neutron richness of baryonic matter and
 the production of light and heavy nuclei.

Experiments with solar, atmospheric and reactor neutrinos have
provided evidence on the oscillations between three neutrino flavors
caused by non-zero neutrino masses
\cite{Esteban:2019,Desalas:2018,Nunokawa:2000,Anselmann:1994, Hampel:1997, Abdurashitov:1996, Abdurashitov:1998}. 

{Different collaborations as the short-baseline neutrinos oscillation experiments LSND (Liquid Scintillator Neutrino Detector) \cite{Athanassopoulos:1996} and MiniBoone (Mini Booster Neutrino Experiment)\cite{Aguilar-Arevalo:2018}, reactor experiments \cite{Mention:2011} and Gallium detectors \cite{Giunti:2011, Acero:2007} have published results presenting anomalies in its collected data which suggest the possible existence of at least one extra neutrino species the so-called sterile neutrino ($\nu_s$),
\cite{Aguilar-Arevalo:2018,Conrad:2013,Aguilar:2007,Kopp:2013,Athanassopoulos:1996,Himmel:2015,Giunti:2011}.}
which is invisible to Z-boson decay and interacts
exclusively in a gravitational way. Data suggest a mass-square
difference $\Delta m^2_{14} \sim 1.3 \textrm{ eV}^2$ and a mixing amplitude $0.1 \leq |{U_{e4}}|\leq 0.22$ \cite{Conrad:2013,Himmel:2015,Giunti:2011,Mention:2011,Conrad:2012,Maltoni:2007}.
More recently, the KATRIN collaboration has reported on the design of the TRISTAN module aimed at the detection of keV sterile neutrinos \cite{houdy:2020}. 

Motivated by these experimental results, the consequences
of the inclusion of sterile neutrinos in different astrophysical scenarios are being
examined. We are interested in the effects due to neutrino
oscillations upon rapid neutron capture processes (r-process)
that produce unstable nuclei which rapidly decay (mostly by $\beta$-decays) to
reach stable isotopes.

Astrophysical data suggest that the formation of nuclei should be
related to processes which took place at earlier stages of the
evolution of the Universe and that continue to occur at present
\cite{Duan:2010}. 
Complementary astrophysical environments for the production of heavy mass nuclei include neutron-star mergers (NSm) and core-collapse supernovae (SN) \cite{Curtis:2018,Cowan:2019,Kajino:2016,Wehmeyer:2015}, { among other processes \cite{Kajino:2019,Thielemann:2011}}.
 The ejecta from NSm have attracted attention after the recent detection
from LIGO collaboration \cite{Kasen:2017,Wu:2019,Watson:2019}.

This work is organized as follows. In Section \ref{formalismo} we
present the formalism needed to compute the heavy nuclear abundances
with the inclusion of the mixing between neutrino mass eigenstates entering the neutrino flavor active sector and an extra sterile neutrino, in two
scenarios: core-collapse supernova and neutron-star merger. In
Section \ref{resultados} we present our results and in Section
\ref{conclusiones} the conclusions are drawn.

\section{Formalism}
\label{formalismo}

The evolution of nuclear abundances is obtained by solving a set of
coupled equations that include different processes; such as, neutron
capture, $\beta$-decay, fission, $\alpha$-decay, $\alpha$-capture,
$\beta$-delayed, neutron emission, photo-dissociation, among others.
In this Section we describe the formalism of $\beta$-decay rates
with the inclusion of one $\nu_s$, and then 
present the details related to the calculation of the nuclear abundances.

\subsection{$\beta$-decay rates including neutrino oscillations}

We start by  writing the Hamiltonian density as { mentioned} in
\cite{Marshak:1969} and \cite{Blin-Stoyle:1973}, that is
\begin{equation}\label{hamil}
\mathcal{H}_\beta=\frac{G_F}{\sqrt{2}}V_{ud}\left(J_\mu L^{\dagger^\mu}+hc \right)\, \, \, ,
\end{equation}
where $G_F$ is the Fermi
coupling constant, $V_{ud}=\cos\theta_{\rm Cabibbo}=0.9738\pm 0.0005$ \cite{yao:2006}. $J_\mu$ and $L_\mu$ are
 the hadronic and leptonic currents, respectively. The transition amplitude is written as
\begin{equation}\label{abeta}
A_{\beta}(n\rightarrow p+e^-+\bar{\nu_e})=\langle p \,\, e^-\,\, \bar{\nu_e}| \int{d^4x}\mathcal{H_{\beta}} |n \rangle \, \, \, .
\end{equation}
In order to obtain the rate we need to compute
$\left|A_{\beta}\right|^2$. The electron-neutrino can be written as
$|\nu_e \rangle =\sum_j U_ {ej} |\nu_j\rangle$, where $U_{ij}$ is
the Pontecorvo-Maki-Nakagawa-Sakata { (PMNS)} mixing matrix \cite{Maki:1962,Giganti:2018} and $|\nu_j\rangle$ is the $j$-mass eigenstates. { In order to reduce the number of parameters entering the calculations, we have set Dirac's and Majorana CP-violating phases at zero  \cite{Fogli:2012,GonzalezGarcia:2012}.}
Furthermore, and in order to perform the calculations, we follow
\cite{Ivanov:2008} and consider a Gaussian package with a radial
spreading $\delta$, that is
\begin{eqnarray}
\label{ivanov1}
\bra{\bar{\nu_j}}{\psi}_{\nu_j}(x,t)&=& \bra{0}\sqrt{\frac{1}{2 q_{\nu_j}^0}} \int\frac{d^3\vec{q}_{\nu_e}}{(2\pi)^3} e^{-(\vec{q}_{\nu_e}-\vec{q}_{\nu_j})^2\frac{\delta^2}{2}} \,e^{-\imath \vec{q}_{\nu_e} \cdot \vec{r}+ \imath q_{\nu_j}^0 t} \nonumber \\
&& \hskip 2cm \times \mathcal{V}_{\nu_j}(q_{\nu_j},s_j)\, \, \, ,
\end{eqnarray}
where ${\psi}_{\nu_j}$ is the field operator of the $j$ particle, $\vec{q}_{\nu_e} \, \, \left(\vec{q}_{\nu_j}\right)$ is the spatial component of the tetra-momentum of the electron-neutrino ($j$-mass-eigenstate) \cite{Cahn:2013}, $\mathcal{V}_{\nu_j}$ is the anti-neutrino Dirac spinor and $q_{\nu_j}^0=\sqrt{|\vec{q}_{\nu_e}|^2+m_j^2}$ is the $j$-neutrino energy, and $m_j$ its mass (all quantities given in natural units $\hbar=c=1$).

The final expression for the decay-rate is obtained after we write
the currents of Eq.(\ref{hamil}) in terms of the particle and
antiparticle fields, make all needed summations on Lorentz and
spin-indexes, and integrate on their spatial momentum. If we call $D_j$ to 
\begin{equation}
D_j=\frac{|U_{ej}|^2 G^2_f |V_{ud}|^2}{(2\pi m_n)^3 }\,,\nonumber
\end{equation}
neglect non-diagonal contributions and take the limit $\delta \rightarrow
0$, we obtain \footnote{There is a misprint in the definition of the factor $D_j$ given in ref \cite{Saez:2020}. Therein $m_n \rightarrow m_n^3$ in the denominator, and $\xi_j \rightarrow \xi_j^2$ in the expression of $\Gamma_{\beta}^0$  given in the same reference.  } 
\begin{eqnarray}
\label{rateconosc}
\Gamma^O_{\beta} &=&\sum_j D_j \int_{a}^{b} dx\sqrt{1-2\frac{\mu_j}{x}+\frac{\xi^2_j}{x^2}} \sqrt{(M-x)^2-4m_p^2 m_n^2}\nonumber\\
&&\hskip 1.25cm\times\Bigg\{\Bigg[\frac{1}{6}(M-x) x \left(1-2\frac{\mu_j}{x}+\frac{\xi^2_j}{x^2}\right)  \nonumber \\
&& \hskip 1.9cm +\frac{2}{3} \left(\frac{M}{2}(M-x)-\frac{(M-x)^2}{4}-m_p^2 m_n^2\right) \nonumber\\
&&\hskip 2.4cm\times\left(1+\frac{\mu_j}{x}-2\frac{\xi^2_j}{x^2}\right)\Bigg] (1+\lambda^2)  \nonumber\\
&& \hskip 1.5cm-(1-\lambda^2)m_n m_p(x-\mu_j)\Bigg\} \, \, \, .
\end{eqnarray}
In the previous expression  $a=(m_e+m_{\nu_j})^2 $, $b=(m_n-m_p)^2$,
$M=m_n^2+m_p^2$, $\mu_j=m_e^2+m_{\nu_j}^2$,
$\xi_j=m_e^2-m_{\nu_j}^2$, $x=M-2q_p^0m_n$, and $\lambda$ is the
weak coupling constant defined as the ratio of the axial vector and
vector coupling constant \cite{yao:2006}. The calculated neutron-decay-rate depends strongly on the mixing
angle $\theta_{14}$, and weakly on the mass-square difference
$\Delta m^2_{14}$ (see \cite{Saez:2020} for details). We worked in the $2+1$
scheme and assumed normal-mass-hierarchy for the neutrino-mass
eigenstates \cite{Meregaglia:2016}, therefore the { PMNS mixing matrix takes the form}
\begin{eqnarray}\label{PMNS}
U&= &
\left(
\begin{array}{ccc}
c_{13}c_{14} & s_{13}c_{14} & s_{14} \\
-s_{13} &c_{13} & 0 \\
-c_{13}s_{14} & -s_{13}s_{14} & c_{14}
\end{array}
\right) \,  \, \, ,
\end{eqnarray}
where $c_{ij}=\cos \theta_{ij}$ and $s_{ij}=\sin \theta_{ij}$. The
oscillation parameters for active-neutrino mixing were taken from
\cite{Meregaglia:2016, Tamborra:2012} and
\cite{Cahn:2013}, that is $\sin^2(2\theta_{13})=0.09$ and $\Delta
m_{13}^2=2\times 10^{-3} \, {\rm eV}^2$.  The parameters for
the sterile neutrino sector $\theta_{14}$ (mixing angle)
and $\Delta m_{14}^2$ (difference between the squared mass
values of the lightest neutrino mass eigenstate and the sterile one) are allowed to vary.
Comparing with the available neutron lifetime measurements of \cite{Tanabashi:2018}, we found that the calculated rates remain within the accepted ranges for $\theta_{14}\leq \pi/9$.

Since we are interested in the impact of oscillations upon the
r-process abundances, we have recalculated the $\beta$-decay rates
for 8055 nuclei involved in the r-process nuclear network used as
input of the differential equations that determine the nuclear
abundances. Therefore, we use the expression given in Eq.
(\ref{rateconosc}) to compute the nuclear rates with the inclusion
of neutrino oscillations $\left( \Gamma_{\beta-nuc}^O\right)$ as
\begin{equation}\label{brates}
\Gamma_{\beta-nuc}^O=\frac{\Gamma_\beta^O}{\Gamma_\beta^{NO}} \Gamma_{\beta-nuc}^{NO} \, \, \, ,
\end{equation}
where $\Gamma_{\beta}^{NO}$ is the $\beta$-decay rate without
including neutrino oscillations.
\subsection{Calculation of heavy nuclei abundances}

{ As mentioned in the Introduction, neutron-star mergers (NSm) and core-collapse supernovae (SN) are the preferred scenarios for producing heavy elements. In Table 1, a set of the observational abundances and their sources is given for medium-heavy nuclear masses in the range $120<\rm{A}<200$. The results are taken from \cite{lodders:2009,roederer:2012}. There is a clear concentration of the abundances ($Y$) of nuclei around the double magic closure $\rm{Z}=50$, $\rm{N}=82$, that is around the tin (Sn) region, and around the next double magic closure $\rm{Z}=82$, $\rm{N}=126$, that is around the lead (Pb) region. However, the fraction of abundances corresponding to deformed nuclei (Eu,Tb,Ho,Tm,Lu,Re) add to approximately $0.40$, which is non-negligible. }

To perform the calculations we have chosen  the r-java2.0 code, developed by the Calgary University group
\cite{charignon:2011,kostka:2014-astr,kostka:2014-nuc}. This code solves the
nuclear r-process network coupled to an astrophysical scenario.  We have modified the inputs of the code to include the $\beta$-decay
rates of Eq.(\ref{brates}), and the set of equations was solved for two
different astrophysical environments:
\begin{enumerate}
\item[a)] {\it a neutrino driven wind in a core-collapse supernova (SN);}

To characterize the neutrino wind, we have chosen the initial temperature as $T_0=3 \times 10^9 \, {\rm K}$, the density profile $\rho=\frac{{\rho}_0}{ (1+t/(2\tau))^2}$ where the initial density is $\rho_0=10^{11} \, {\rm g/cm}^3$ and the expansion time-scale is $\tau=0.1 \, {\rm s}$. The wind speed of expansion was set to $V_{\rm exp}=7500 \, {\rm km/s}$ and the initial wind radius fixed at $R_0=390 \, {\rm km}$ \cite{kostka:2014-astr,charignon:2011,meyer:1992,meyer:1997,Tamborra:2012,Wu:2013}.

\item[b)] {\it a neutron-star merger (NSm);}

This environment was characterized by an initial temperature $T_0=3 \times 10^9 \, {\rm K}$ and the polytropic profile $P=K{\rho}^{(n+1)/n}$ where the polytropic index is fixed at $n=1.5$ and $K$ is constant \cite{kostka:2014-astr}. The initial density and the internal pressure were set to $\rho_0=10^{11} \, {\rm g/cm}^3$ and $P_0=0.005 \, {\rm MeV/fm}^3$ respectively. Finally, the expansion velocity was fixed at $V_{exp}=1.9 \times 10^5 \, {\rm km/s}$ and the initial radius of the ejecta at $R_0= 2\, {\rm km}$ \cite{charignon:2011,oechslin:2007}.

\end{enumerate}

In both environments, we have considered two different initial seed distribution
\begin{itemize}
\item[i)] a distribution of mass-fractions corresponding to an entropy of $2.36$ ($s_{100}$) \cite{farouqi:2010};
\item[ii)] a nuclear statistic equilibrium distribution (NSE) \cite{arnould:2007,qian:2003}.
\end{itemize}

For the case ii) we have used, in the core-collapse supernova
environment, an initial electron fraction $Y_e=0.3$. For the
neutron-star mergers we have considered two values for $Y_e$, namely
$0.1$ and $0.3$, and performed a combination of both computed
abundances to obtain their final values, as done in
\cite{arnould:2007,qian:2003,kostka:2014-astr}.
\section{Results and Discussions}
\label{resultados}
{In this section, we shall describe details on the calculations, and advance the discussion for two scenarios:}

\begin{itemize}
\item Case I: abundances generated when only $\beta$-decay and n-captures are included in the network;
\item Case II: { the same of Case I with the addition of fission channels.}
\end{itemize}

In Cases I and II, both for SN and NSm, we have studied
the abundances for different decay-times, $\left(t_{sta}\right)$,
towards stability. We have observed that, for $\left(t_{sta}=0 \,
{\rm s}\right)$, a large abundance of heavy elements with $\rm{A}>200$ is
found. If $t_{sta}$ relaxes to a value of the order of the age of
the Universe, the heaviest nuclei (masses $\left(\rm{A}>200\right)$),
decay to $^{232}$Th, $^{235}$U and $^{238}$U. We have adopted, for
the stability time, the value $t_{sta}= 4.354 \times 10^{17} \, {\rm
s}$ \cite{kostka:2014-astr,charignon:2011,arnould:2007}.

Since in the calculations only the production of nuclei through the
r-process is considered, we have restricted the observational data
\cite{lodders:2009} to the abundances produced primarily through
it. \cite{Bisterzo:2011,Sneden:2003,roederer:2012,sneden:1996}. The
observational data are shown in Table 1. The data and
the results given by the numerical code are scaled by a factor $F$
\begin{eqnarray}
Y_{th}&=& F Y_{code} \, \, \, ,
\label{th-data}
\end{eqnarray}
where $Y_{th}$ is the abundance compatible with the data, and
$Y_{code}$ is the calculated value. The factor $F$ depends on the initial mass-fraction distribution used and of the environment \cite{zhang:2010,arnould:2007,charignon:2011,lanfranchi:2008,cowley:1980}.
 {The r-java code \cite{kostka:2014-astr,kostka:2014-nuc,charignon:2011} requires a large number of variables as input, namely: astrophysical and nuclear variables like $T_0$ (the initial temperature), $\rho_0$ (density profile), $V_{exp}$ (expansion velocity), $R_0$ (initial radii),  among others. Each of the calculations gives different values for the scaling factor $F$ of Eq. (\ref{th-data}). Table 2 shows the values of $F$ which better relate theoretical and observational abundances (see Eq. (\ref{th-data})), for the adopted initial conditions, like the electron fractions $Y_e$ and the entropy $s_{100}$.}

\begin{table}[pt] 
\tbl{ The observational abundances {($Y$) extracted from different sources. The mass (A), charge (Z), and the name of the element (E), are listed in the first three columns for each nucleus. The fourth column indicates the source of the data: average of solar and meteoritic values (a); meteoritic based on CI-chondrites (m);  values extracted from theoretical models (t) \cite{lodders:2009,roederer:2012}. $\sigma$ denotes the standard deviation of the abundances.}}
{\begin{tabular}{@{}ccccc@{}} \toprule
A & Z & E & Source & $Y \pm \sigma$\\
\colrule
127&53&I&m&$1.10 \pm 0.22$\\
130&52&Te&m&$4.69 \pm 0.33$\\
132&54&Xe&t&$5.46 \pm 1.10$\\
153&63&Eu&a&$0.0984 \pm 0.0106$\\
159&65&Tb&m&$0.0634 \pm 0.0044$\\
165&67&Ho&m&$0.0910 \pm 0.0064$\\
169&69&Tm&m&$0.0406 \pm 0.0028$\\
175&71&Lu&m&$0.0380 \pm 0.0019$\\
187&75&Re&m&$0.0554 \pm 0.0055$\\
192&76&Os&m&$0.680 \pm 0.054$\\
193&77&Ir&a&$0.672 \pm 0.092$\\
195&78&Pt&m&$1.27 \pm 0.10$\\
197&79&Au&m&$0.195 \pm 0.019$\\
\botrule
\end{tabular}}
\end{table}

\begin{table}[pt]
\tbl{Scaling factor $F$ of Eq. (\ref{th-data}). The  environment and initial conditions corresponding to the calculations are indicated in the first and second column, respectively. $s_{100}$ and $Y_e$ stand for the entropy and the initial electron fraction.}
{\begin{tabular}{@{}ccc@{}} \toprule
Environment & Initial condition & $F$ \\
\colrule
SN &$s_{100}$&$1564$ \\
SN & NSE $\left(Y_e=0.3\right)$ & $2790$ \\
NSm &$s_{100}$&$3097$ \\
NSm &NSE $\left(Y_e=0.1\right)$ & $3196$\\
NSm &NSE $\left(Y_e=0.3\right)$ & $5709$ \\
\botrule
\end{tabular}}
\end{table}
{To start with, in order to figure out about the impact of the SN and NSm scenarios in the more conventional way, we have computed the abundances without taking into account neutrino oscillations. }

\begin{figure}[!ht]
\epsfig{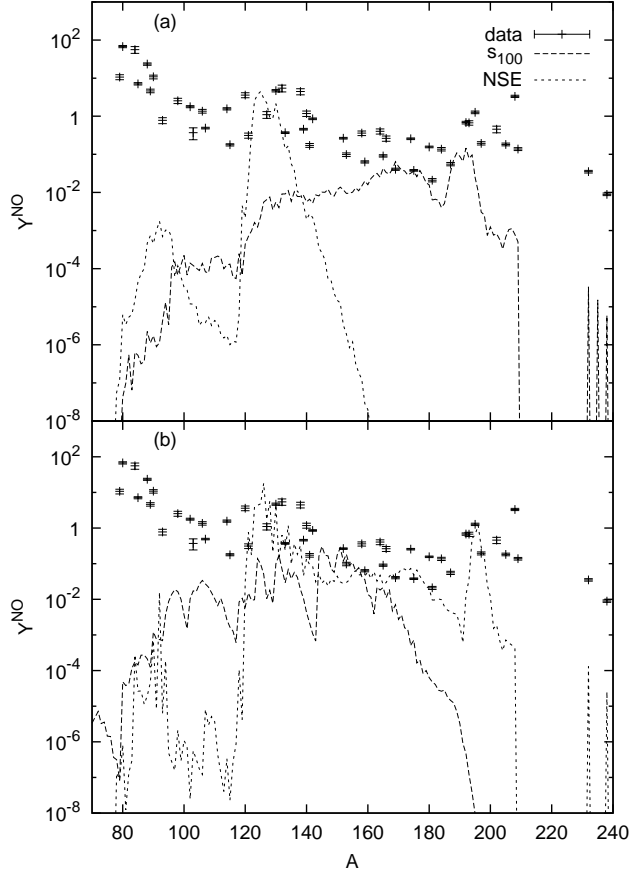}
\caption{Calculated nuclear abundances without the { inclusion} of neutrino
oscillations ($Y^{NO}$) as a function of the mass number (A) for
Case I (without fission). {The $s_{100}$ mass fraction distribution is denoted by a dashed-line; dotted-line represents the NSE initial mass fraction distribution and the dots with error bars are the observational data \cite{lodders:2009}. Insets (a) and (b) show the results with SN and NSm environments, respectively.}} \label{abundancias-A}
\end{figure}

\begin{figure}[!ht]
\epsfig{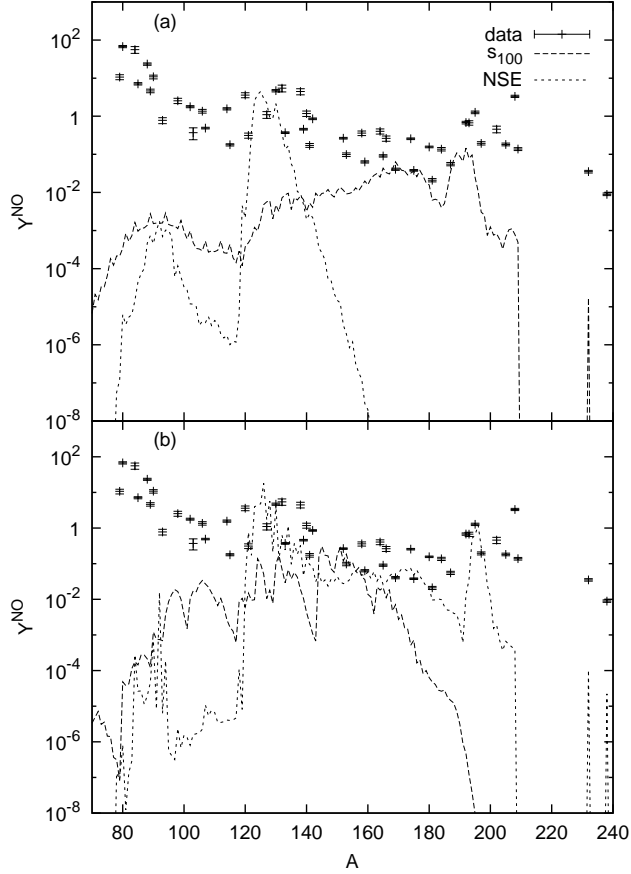}
\caption{Calculated nuclear abundances without the inclusion of neutrino
oscillations ($Y^{NO}$) as a function of the mass number (A) for
Case II.  The $s_{100}$ mass fraction distribution is denoted by a dashed-line;  dotted-line represents the NSE initial mass fraction distribution and the dots with error bars are the observational data \cite{lodders:2009}. Insets (a) and (b) show the results with SN and NSm environments, respectively.} \label{abundancias-B}
\end{figure}

Figures \ref{abundancias-A} and \ref{abundancias-B} show the nuclear
abundances obtained without including neutrino oscillations
$(Y^{NO})$ as a function of the mass number A, for Cases I and II,
respectively. These abundances were corrected by the scaling factor
$F$ and insets (a) and (b) of each figure correspond to the SN (inset (a)) and NSm
(inset (b)) environments.

As we can see in Figure \ref{abundancias-A} (Case I), the abundances calculated with the initial condition set as NSE have two maxima, one at $\rm{A}\sim 80$ and the other at $\rm{A}\sim 130${, being the result in the region $120<\rm{A}<140$ close to the data}. For the NSm scenario,
there appears a third peak located at $\rm{A}\sim 195$. For the $s_{100}$ initial condition, the position of the peak depends on the environment, that is $\rm{A} \sim 190$ for core-collapse supernova (inset (a)) and $\rm{A} \sim 150$ for neutron-star merger (inset (b)). {From the results shown in Figure \ref{abundancias-A}, we may conclude that the agreement with data is rather poor for the SN environment for both initial conditions (NSE and $s_{100}$) except for few points in region $\rm{A}>120$. NSm environment gives much better results. Although the production of light-mass elements is grossly underestimated in both scenarios, something which is expected because the production of these elements can not be attributed only to r-process.}

The results  obtained with the inclusion of fission, see Figure \ref{abundancias-B}, are similar to those of Figure \ref{abundancias-A}.

{ Since we are considering only SN and NSm environments \footnote{See ref. \cite{Kajino:2019,Thielemann:2011} for processes other than SN and NSm.}, to have an idea about their relative effect upon the observed abundances, we summed up the contribution to the abundances of each environment, by expressing them as the linear combination}
\begin{eqnarray}
Y_{\omega}&=& \omega Y_{th\, SN} + \left(1-\omega\right) Y_{th\, NSm} \, \, \,  .
\label{yw}
\end{eqnarray}
{ The factor $\omega$ varies in the interval $0 \leq \omega \leq 1 $, $Y_{th\, SN}$ and $Y_{th\, NSm}$ are
the calculated abundances given by the SN and NSm
environments, respectively.
To extract the actual value of $\omega$, we have taken the observed abundances of Table 1 and compared them with the theoretical results obtained with the set of parameter given in Section 2.2. This procedure yields   $w=0.32$ with a standard deviation of $\pm 0.04$,}
{ a result which suggests that the NSm mechanism dominates over the SN, at least for this picture where only this two environments are considered. The results obtained so far in this Section, do not take the neutrino oscillations into account.} 

\subsection{Impact of neutrino oscillations upon the heavy nuclei abundances}

Hereon we shall { include neutrino oscillations in the calculation of the abundances. The mixing with sterile neutrinos is accounted by the matrix $U$ of  Eq. (\ref{PMNS})}. 

\begin{figure}[!ht]
\epsfig{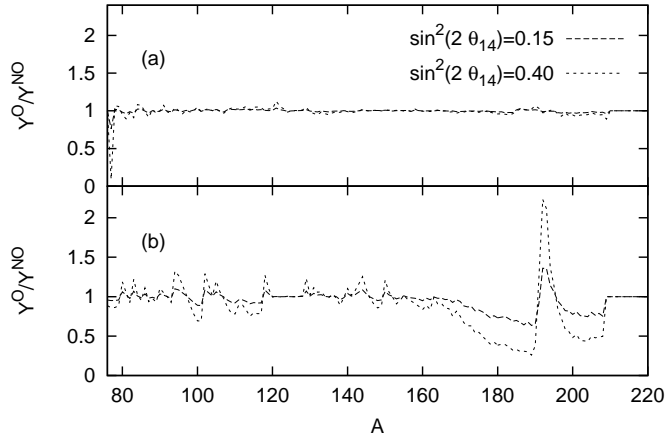} \caption{Ratio
between { calculated} abundances with { ($Y^{O}$)} and without {($Y^{NO}$)} neutrino oscillations, as a function of the mass number $\rm{A}$ for Case I. { The neutrino oscillation parameters are the ones listed in the text, with} $\Delta m^2_{14}=1 \, {\rm eV}^2$. Dashed-line: $\sin^2(2\theta_{14})=0.15$, dotted-line:
$\sin^2(2\theta_{14})=0.40$. Inset (a): SN environment; inset (b):
NSm environment.} \label{oscilaciones-A}
\end{figure}
\begin{figure}[!ht]
\epsfig{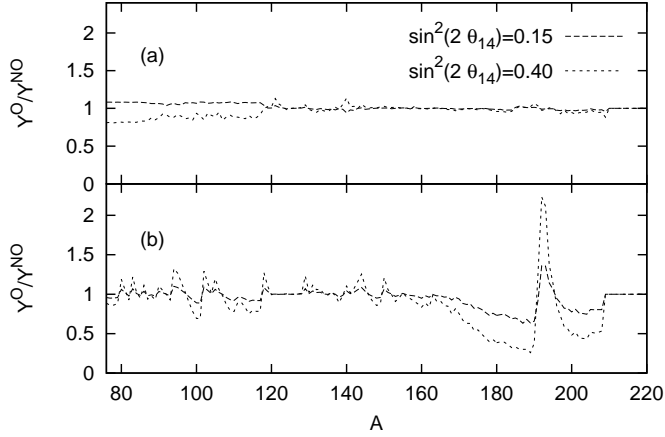}
\caption{Ratio between abundances with and without neutrino oscillations, $Y^{O}/Y^{NO}$, as a function of the mass number $\rm{A}$ for Case II. { The oscillation parameters are those given in the caption of Figure \ref{oscilaciones-A}.} Dashed-line: $\sin^2(2\theta_{14})=0.15$, dotted-line: $\sin^2(2\theta_{14})=0.40$. Inset (a): SN environment; inset (b): NSm environment.}
\label{oscilaciones-B}
\end{figure}

Figures \ref{oscilaciones-A} and \ref{oscilaciones-B} illustrate the effects { of the neutrino oscillations}, for Cases I and II, respectively. The inclusion of neutrino oscillations changes the relative $\beta$-decay rates along the r-process path and it affects the {heavy-nuclei} abundances. The results have been obtained { by varying the values of the mixing-angle $\theta_{14}$ and of the square-mass difference $\Delta m^2_{14}$},{  and taken } $\sin^2(2\theta_{13})=0.09$ and $\Delta m^2_{13}=2\times 10^{-3} \,
{\rm eV}^2$ \cite{Tamborra:2012,Balantekin:2005,Cahn:2013,Meregaglia:2016}. { For the sterile
neutrino sector in both figures we have taken the values } $\Delta m^2_{14}=1 \, {\rm eV}^2$ and $\sin^2(2\theta_{14})=0.15$ (dashed-line) and
$\sin^2(2\theta_{14})=0.40$ (dotted-line). {Also, we have performed additional calculations in a wider parameter space ($0\leq \sin^2{2\theta_{14}}\leq 1$ and  $0.5 \, \rm{eV}^2 \leq \Delta m^2_{14} \leq 2.5 \, \rm{eV}^2$) to account for other values for the anomaly fitting statistics \cite{Boser:2020,Dentler:2018,Gariazzo:2015,Diaz:2019}.  The results are not sensitive to changes in $\Delta m^2_{14} $.}

{ In both Figures \ref{oscilaciones-A} and \ref{oscilaciones-B}  (see Inset (a))}, the effects of neutrino
oscillations in the SN environment { become unnoticed, while for the NSm environment (Inset (b)), the ratio $Y^O/Y^{NO}$ varies substantially for heavy-mass elements. This behavior may be attributed to the changes in the $\beta$-decay rates along the r-process path}.

It is seen, Figure \ref{oscilaciones-A}, that for the SN environment, it exists an
underproduction of $^{77}{\rm Se}$  respect to the case without neutrino oscillations. In the same Figure \ref{oscilaciones-A}, for the case of NSm, there  is also a  light overproduction of certain nuclei
$\left(\rm{A}=94, \, 95, \, 102, \, 118, \, 144\right)$. 
{ The gross mechanism underlying this effect is related to changes induced on the weak decay rates, which is not the same all along different mass regions. }
We have noticed that the inclusion of sterile neutrinos in the calculations particularly affects the third r-process peak region,
lowering the $\beta$-decay rate of elements like Dy, Ho, Er, Tm, Yb, Ir, Os, yielding higher abundances around $\rm{A}=192$ and $\rm{A}=195$ $\left(^{192}\rm Os \,\, \rm{and}\,\, ^{195}\rm Pt\right)$, as it is shown in inset (b) of Figure \ref{oscilaciones-A}.
In contrast, there is also an underproduction of elements with $156<\rm{A}<190$ and $196<\rm{A}<208$ if the active-sterile neutrino oscillations are taken into account.

For  Case II, consisting of the inclusion of fission in the network and large
active-sterile neutrino mixing, we observe an underproduction
of elements with $76<\rm{A}<118$ in the core-collapse supernova environment (inset (a)
of Figure \ref{oscilaciones-B}). For the neutron-star merger scenario (inset (b) of Figure \ref{oscilaciones-B}), { the inclusion of massive sterile neutrinos in the formalism give similar results as for Case I.} 
The effects produced by the mixing with $\nu_s$ seem to be larger
in the NSm environment{. Further comments on the significance of the values of the mixing angle $\theta_{14}$ are given in the next subsection. }

\subsection{Constraints on the mixing parameters of the neutrino sector}
{As mentioned in the Introduction there is a growing piece of evidence about the existence of sterile neutrino. The values for the mixing angle $\theta_{14}$ and the square-mass difference $\Delta m_{14}^2$ which we have considered so far have been taken from \cite{Conrad:2013,Himmel:2015,Giunti:2011,Mention:2011,Conrad:2012,Maltoni:2007}. By other hand we can reverse the argument and from the comparison between observed and calculated nuclear abundances we may extract values on the same quantities.

By performing an statistical analysis \footnote{Consisting of extracting the average value that best reconciles the calculated abundances with the data over the 520 runs.} using the observed abundances, values of $Y$ given in Table 1, and the calculated ones for different values of the mixing angles and for a fixed square mass difference $\Delta m_{14}^2= 1 \rm{eV}^2$ we arrive at the value}

\begin{eqnarray}\label{best-fit-values}
\sin^2(2\theta_{14}) &=& 0.22^{+0.13}_{-0.15} 
\end{eqnarray}

{ A value which is not in tension with the one extracted by other means \cite{Aguilar-Arevalo:2018,Conrad:2013,Aguilar:2007,Kopp:2013,Athanassopoulos:1996,Himmel:2015,Giunti:2011}. }

\begin{figure}[!ht]
\epsfig{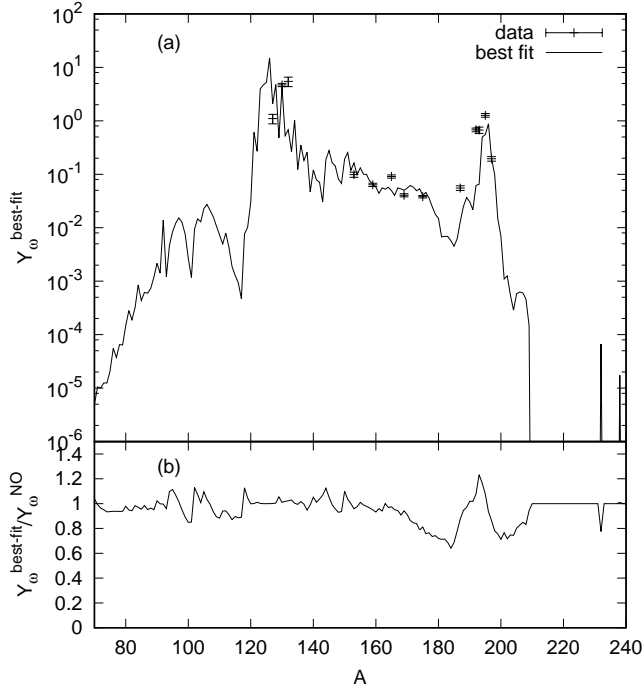} \caption{ Inset (a): nuclear abundances calculated with the pair of central values ($\sin^2{2\theta_{14}}=0.22, \omega=0.32$) as a function of the mass number ($\rm{A}$) for
Case II (solid-line), compared with data \cite{lodders:2009} (dots with error bars). Inset (b): ratio between the abundances with and without neutrino oscillations $Y^{best-fit}_\omega/Y^{NO}_\omega$, calculated with the pair of central values  for the same Case II.}
\label{abundancia-ajuste}
\end{figure}

{ The inset (a) of Figure \ref{abundancia-ajuste} shows the nuclear
abundances as a function of the mass number $(\rm{A})$ for the Case II
calculated  with the value of $\sin^2(2\theta_{14})$ given in Eq. (9). }

The ratio between the abundances of Eq. (8), calculated { with and without } neutrino mixing with $\sin^2(2\theta_{14})=0.22$ and $\omega=0.32$ is shown in the inset (b) of Figure \ref{abundancia-ajuste}. The agreement between calculations and data (Figure \ref{abundancia-ajuste} (a)) is rather good. { It is noted that the trend of the theoretical results around $^{195}\rm Pt$ follows the trend of the data. As a matter of fact the calculated values go up for $190\leq \rm{A} \leq 195$ and down for $\rm{A}>195$, a feature which agrees with the data}. Concerning the ratio $Y^{best-fit}_\omega/Y^{NO}_\omega$ (Figure \ref{abundancia-ajuste} inset (b)) the larger effects are shown for masses between $180 \leq \rm{A} \leq 220$ but the overall departure between both predictions amount to 20\%, except for the just mention fluctuation. {The abundances of elements like  Te, Eu, Tb, Tm, Lu, and Au are reasonable reproduced while the abundances of Ho, Xe, Re, Os, and Ir are underpredicted, although the overall tendency of the calculation follows the data as mentioned before. It is expected that the agreement between data and predictions could improve if: i) a better acknowledgment of the r-process is supply by the experiments; ii) by improving the nuclear network code; iii) by collecting more information about the SN and NSm scenarios; iv) by taking other initial conditions and/or scenarios.}

\section{Conclusions}
\label{conclusiones}

As a first step we have analyzed the contribution to the heavy mass
abundances for each of the NSm and SN environments { without including neutrino oscillations}. 

{ Then, we have included neutrino oscillations in the calculation of the decay rates and found that they affect the heavy nuclear abundances as well. The effect }on the calculated
neutron-decay-rate is larger for larger values of the mixing angle $\theta_{14}$. For the SN environment the major effects due to active-sterile
neutrino mixing reflect upon the domain $76 < \rm{A} < 118$. For the NSm environment, the effects are larger for Zr, Mo,
Ru, Sn, Nd, and Os $\left(\rm{A}=94, \, 95, \, 102, \, 118, \, 144, \,
192\right)$.

{ We have performed a statistical test to set limits on the mixing
angle between active and sterile neutrinos using observational data
\cite{lodders:2009} and found the central values
$\sin^2 2\theta_{14}=0.22$. Since the result corresponding to the mixing with sterile neutrinos seem to be better than those obtained in the absence of oscillations, at least for the two considered environments, a systematic study of the abundances of nuclear species produced in astrophysical environments may be a source of information about sterile neutrinos other than direct experiments. Concerning the dominance of NSm respect to SN environment found in the calculations, represented by the factor $\omega$, we would like to emphasize that it can be taken as indicative since only this two scenarios have been considered. }
We hope that future and complementary studies will help to reconcile the observations with the predictions, as well as to provide a definite conclusion about the existence of eV-scale sterile neutrinos.

\section*{{ Acknowledgment}}
This work was supported by a grant (PIP-616) of the National Research Council of Argentina (CONICET), and by a research-grant
(PICT No. 140492 ) of the National Agency for the Promotion of Science and Technology (ANPCYT) of Argentina. O. C. and M. E. M. are members of the Scientific Research Career of the CONICET, M. M. S. is a Post Doctoral fellow of the CONICET and K. J. F is a Ph.D fellow of the CONICET.


\begin{thebibliography}{0}
\bibitem{Esteban:2019} I. Esteban, {Journal of High Energy Physics} {\bf 1}, 106 (2019).
\bibitem{Desalas:2018} P.F. de Salas, D.V. Forero, C.A. Ternes, M. T\'ortola and J.W.F. Valle, Physics Letters B {\bf 782}, 633 (2018).
\bibitem{Nunokawa:2000} H. Nunokawa,Brazilian Journal of Physics {\bf 30}, 346 (2000). 

\bibitem{Anselmann:1994}
P.~Anselmann \textit{et al.} [GALLEX],
Phys. Lett. B \textbf{342}, 440-450 (1995)
doi:10.1016/0370-2693(94)01586-2
\bibitem{Hampel:1997}
W.~Hampel \textit{et al.} [GALLEX],
Phys. Lett. B \textbf{420}, 114-126 (1998)
doi:10.1016/S0370-2693(97)01562-1

\bibitem{Abdurashitov:1996}
D.~N.~Abdurashitov, \textit{et al.} [SAGE]
Phys. Rev. Lett. \textbf{77}, 4708-4711 (1996)
doi:10.1103/PhysRevLett.77.4708



\bibitem{Abdurashitov:1998}
J.~N.~Abdurashitov \textit{et. al.} [SAGE],
Phys. Rev. C \textbf{59}, 2246-2263 (1999)
doi:10.1103/PhysRevC.59.2246
[arXiv:hep-ph/9803418 [hep-ph]].

\bibitem{Athanassopoulos:1996} C. Athanassopoulos et al., Phys. Rev. Lett. {\bf 77}, 3082 (1996).
\bibitem{Aguilar-Arevalo:2018} A. A. Aguilar-Arevalo et al., Phys. Rev. Lett. {\bf 121}, 221801 (2018).

\bibitem{Mention:2011}
G.~Mention, M.~Fechner, T.~Lasserre, T.~A.~Mueller, D.~Lhuillier, M.~Cribier and A.~Letourneau,
Phys. Rev. D \textbf{83}, 073006 (2011)
doi:10.1103/PhysRevD.83.073006
[arXiv:1101.2755 [hep-ex]].

\bibitem{Giunti:2011}
C.~Giunti and M.~Laveder,
Phys. Rev. D \textbf{84}, 073008 (2011)
doi:10.1103/PhysRevD.84.073008
[arXiv:1107.1452 [hep-ph]].
\bibitem{Acero:2007}
M.~A.~Acero, C.~Giunti and M.~Laveder,
Phys. Rev. D \textbf{78}, 073009 (2008)
doi:10.1103/PhysRevD.78.073009
[arXiv:0711.4222 [hep-ph]].



\bibitem{Conrad:2013} J. Conrad, W. Louis, and  M. H. Shaevitz, Ann. Rev. Nucl. Part. Sci. {\bf 63}, 5 (2013).
\bibitem{Aguilar:2007} A. A. Aguilar-Arevalo et al. Phys. Rev. Lett. {\bf 98}, 231801 (2007).
\bibitem{Kopp:2013} J. Kopp,  P. A. Machado, M. Maltoni, T. Schwetz, Journal of High Energy Physics {\bf 05} 050, (2013).

\bibitem{Himmel:2015} A. Himmel, Physics Procedia {\bf 61}, 612 (2015).
\bibitem{Conrad:2012} J.M. Conrad, C.M. Ignarra, G. Karagiorgi, M.H. Shaevitz and J. Spitz, arXiv e-prints 1207.4765, (2012).
\bibitem{Maltoni:2007} M. Maltoni, M. and T. Schwetz, Phys. Rev. D.{\bf 76}, 093005 (2007).
\bibitem{houdy:2020} T. Houdy et al., Journal of Physics Conference Series {\bf 1468}, 012177 (2020). 
\bibitem{Duan:2010} H. Duan, A. Friedland, G. McLaughlin, and R. Surman, J. Phys G {\bf 38}, 035201 (2011).
\bibitem{Curtis:2018} S. Curtis et al., The Astrophysical Journal {\bf 870}, 2 (2018).
\bibitem{Cowan:2019} J. J. Cowan et al.arXiv:1901.01410.
\bibitem{Kajino:2016} T. Kajino and G.J. Mathews, Rept. Prog. Phys.{\bf 80}, 084901 (2017).
\bibitem{Wehmeyer:2015} B. Wehmeyer, M. Pignatari and F.K. Thielemann,  Mon. Not. Roy. Astron. Soc. {\bf 452}, 1970 (2015).
\bibitem{Kajino:2019}
T.~Kajino, W.~Aoki, A.~B.~Balantekin, R.~Diehl, M.~A.~Famiano and G.~J.~Mathews,
Prog. Part. Nucl. Phys. \textbf{107}, 109-166 (2019)
doi:10.1016/j.ppnp.2019.02.008
[arXiv:1906.05002 [astro-ph.HE]].

\bibitem{Thielemann:2011} Thielemann, F.-K., Arcones, A., K{\"a}ppeli, R., et al.\  Progress in Particle and Nuclear Physics, {\bf 66}, 346 (2011).



\bibitem{Kasen:2017} D. Kasen, B. Metzger, J. Barnes, E. Quataert and E. Ramirez-Ruiz, Nature {\bf 551}, 80(2017).
\bibitem{Wu:2019} M.R. Wu, J. Barnes, G. Martinez-Pinedo and B.D. Metzger, Phys. Rev. Lett. {\bf 122}, 062701 (2019).
\bibitem{Watson:2019} D. Watson et al., Nature {\bf 574}, 497 (2019).
\bibitem{Marshak:1969} R. E. Marshak, Riaduzzin and C. P. Ryan, \emph{Theory of weak interactions in particle physics} (Wiley-Interscience; New York, 1969).
\bibitem{Blin-Stoyle:1973} R. J. Blin-Stoyle, Fundamental Interactions and the Nucleus, Ed.North-Holland Publishing Company; New York (1973).
\bibitem{yao:2006} W. M. Yao et al., Journal of Physics G {\bf 33}, 1 (2006).

\bibitem{Maki:1962}
Z.~Maki, M.~Nakagawa and S.~Sakata,
Prog. Theor. Phys. \textbf{28}, 870-880 (1962)
doi:10.1143/PTP.28.870

\bibitem{Giganti:2018}
C.~Giganti, S.~Lavignac and M.~Zito,
Prog. Part. Nucl. Phys. \textbf{98}, 1-54 (2018)
doi:10.1016/j.ppnp.2017.10.001
[arXiv:1710.00715 [hep-ex]].


\bibitem{Fogli:2012}
G.~L.~Fogli, E.~Lisi, A.~Marrone, D.~Montanino, A.~Palazzo and A.~M.~Rotunno,
Phys. Rev. D \textbf{86}, 013012 (2012)

\bibitem{GonzalezGarcia:2012}
M.~C.~Gonzalez-Garcia, M.~Maltoni, J.~Salvado and T.~Schwetz,
JHEP \textbf{12}, 123 (2012)


\bibitem{Ivanov:2008} A.N. Ivanov, R. Reda, and P. Kienle. arXiv:0801.2121.
\bibitem{Cahn:2013} R. N. Cahn et.al., Proceedings of the 2013 Community Summer Study on the Future of U.S. Particle Physics: Snowmass on the Mississippi (CSS2013): Minneapolis, MN, USA, July 29-August 6, (2013), arxiv:1307.5487.
\bibitem{Saez:2020} M.M. Saez, O. Civitarese and M. E. Mosquera, International Journal of Modern Physics E {\bf 29}, 2050022 (2020).
\bibitem{Meregaglia:2016} A. Meregaglia. (The Double Chooz Collaboration), Nuovo Cimento Geophysics Space Physics C, {\bf 38}, 123  (2016).
\bibitem{Tamborra:2012} I. Tamborra, G. G. Raffelt, and D. V Semikoz, JCAP {\bf 1}, 13 (2012).
\bibitem{Tanabashi:2018} M. Tanabashi et al., Phys. Rev. D. {\bf 98}, 030001 (2018).
\bibitem{charignon:2011} C. Charignon, M. Kostka, N. Koning, P. Jaikumar and R. Ouyed, Astron. Astrophys. {\bf 531}, A79 (2011).
\bibitem{kostka:2014-astr} M.~Kostka, N.~Koning, Z.~Shand, R.~Ouyed and P.~Jaikumar (2014), arXiv:1402.3824 [astro-ph.IM].
\bibitem{kostka:2014-nuc}  M.~Kostka, N.~Koning, Z.~Shand, R.~Ouyed and P.~Jaikumar, Astron.\ Astrophys.\  {\bf 568}, A97 (2014).
\bibitem{meyer:1992} B. S. Meyer,  G. J. {Mathews},  W. M.  {Howard}, S. E.{Woosley} and R. D. {Hoffman}, ApJ {\bf 399}, 656 (1992).
\bibitem{meyer:1997} B. S. Meyer and J. S. Brown, The Astrophysical Journal Supplement Series {\bf 112}, 199 (1997).
\bibitem{Wu:2013} M.~R.~Wu, T.~Fischer, L.~Huther, G.~Martínez-Pinedo and Y.~Z.~Qian, Phys.\ Rev.\ D {\bf 89}, no. 6, 061303 (2014).
\bibitem{oechslin:2007} R. Oechslin. H.T. Janka and A. Marek, Astronomy and Astrophysics {\bf 467}, 395 (2007).
\bibitem{farouqi:2010} K. Farouqi, K.L.Kratz, B. Pfeiffer, T. Rauscher, F.K.Thielemann and J.W.Truran,  Astro Particle Physics {\bf 712}, 1359 (2010).
\bibitem{arnould:2007} M. Arnould, S. Goriely and K. Takahashi, Phys. Rep. {\bf 450}, 97 (2007).
\bibitem{qian:2003} Y.-Z. Qian, Prog. Part. Nucl. Phys. {\bf 50}, 153 (2003).
\bibitem{lodders:2009} K. Lodders, H. Palme, H.P. Gail, Landolt Boumlrnstein {\bf 4B}, 712 (2009). 
\bibitem{Bisterzo:2011} S. Bisterzo et al., Mon. Not. Roy. Astron. Soc. {\bf 418}, 284 (2011).
\bibitem{Sneden:2003} C. Sneden et al., The Astrophysical Journal {\bf 591}, 936 (2003). 
\bibitem{roederer:2012} I.U. {Roederer} and J.E. {Lawler}, The Astrophysical Journal {\bf 750}, 76 (2012). 
\bibitem{sneden:1996} C. Sneded et al.,  The Astrophysical Journal {\bf 467}, 819 (1996).
\bibitem{zhang:2010} J. Zhang, W. Cui, Wenyuan and B. Zhang, Mon. Not. Roy. Astron. Soc. {\bf 409}, 1068 (2010).
\bibitem{lanfranchi:2008} G.A. Lanfranchi, F. {Matteucci}, F. and G. {Cescutti}, Astronomy and Astrophysics {\bf 481}, 635 (2008).
\bibitem{cowley:1980} C.R. Cowley and P.L. {Downs}, The Astrophysical Journal {\bf 236}, 648 (1980). 
\bibitem{Balantekin:2005} A.B. Balantekin and H. Yüksel, New Journal of Physics {\bf 7}, 51 (2005).
\bibitem{Boser:2020} S. B{\"o}ser et al., Progress in Particle and Nuclear Physics {\bf 111}, 103736 (2020).
\bibitem{Dentler:2018} M. Dentler et al., Journal of High Energy Physics {\bf 8}, 10 (2018). 
\bibitem{Gariazzo:2015} S. Gariazzo, C. Giunti,  M. Laveder,  Y. F. Li and E. M. Zavanin, Journal of Physics G: Nuclear and Particle Physics {\bf 43}, 033001 (2015). 
\bibitem{Diaz:2019} A. Diaz et al., arXiv e-prints 1906.00045 (2019). 

\end{thebibliography}
\end{document}